\documentclass[reqno, eucal]{amsart}

\usepackage[mathscr]{eucal} %  use \EuScript (\mathcal unchanged)
\usepackage{here}

\usepackage[dvips]{graphicx}
\usepackage[dvips]{color}
\usepackage{amsmath,amsfonts,amssymb,amsthm,amscd}
\usepackage{epic}
\usepackage{eepic}
\usepackage{longtable}
\usepackage{array}

\usepackage{fancyhdr}
\fancypagestyle{foot}
{
\fancyhead{}
\fancyfoot{}
\fancyfoot[C]{\thepage}
}
\newtheorem{theorem}{Theorem}[section]

\theoremstyle{definition}

\newcommand{\Z}{{\mathbb Z}}

\def\red#1{{\color{red}#1}}
\def\blue#1{{\color{blue}#1}}
\begin{document}

\begin{center}
\begin{Large}
On tetrahedron type equations 
associated with $B_3,C_3, F_4$ and $H_3$
\end{Large}
\end{center}

\vspace{0.5cm}
\begin{center}
Atsuo Kuniba\\
\vspace{0.2cm}
Institute of Physics, Graduate School of Arts and Sciences, 

University of Tokyo, Komaba, Tokyo 153-8902 Japan
\end{center}

\vspace{0.5cm}
\begin{center}
Abstract
\end{center}
Tetrahedron equation is a three dimensional analogue of the Yang-Baxter equation.
It allows a formulation in terms of the Coxeter group $A_3$.
This short note includes miscellaneous remarks on the 
generalizations along $B_3, C_3, F_4$ and the non-crystallographic Coxeter group $H_3$.
It is a supplement to the author's talk in the online workshop
{\em Combinatorial Representation Theory and Connections with Related Fields} at 
RIMS, Kyoto University in November 2021.

\vspace{0.5cm}

\section{Introduction}

Yang-Baxter equation \cite{Jim90} plays a central role in solvable lattice models in two dimension \cite{Bax82}  and 
integrable quantum field theories in $1+1$ dimensional space time \cite{Zam79}.
Tetrahedron equation \cite{Zam80} is an analogue of the Yang-Baxter equation in three dimensional space.
It is naturally endowed with the Coxeter group $A_3$ \cite{KV94}. 

One can generalize the equations and solutions  
from the viewpoint of finite Coxeter groups \cite{KO12,KO13,Kun18-1}.
Given a rank $n$ Coxeter group $X_n$, 
a common feature is the correspondence \cite{Ku22}:
\begin{equation}\label{szk1}
\begin{split}
\text{basic operators}  &\leftrightarrow  \text{Coxeter relations in $X_n$},
\\
\text{tetrahedron type equation} & \leftrightarrow
\text{inclusion $X_n \hookrightarrow X_{n+1}$ as a parabolic subgroup}.
\end{split}
\end{equation}

In contrast,
the correspondence 
in two dimension \cite{Ch84} holds between the Yang-Baxter equation and the cubic 
Coxeter relation $s_is_js_i = s_js_is_j$ of the generators, the reflection equation and the 
quartic one $s_is_js_is_j = s_js_is_js_i$ and the $G_2$ reflection equation \cite{Kun18-1} and 
the sextic one $s_is_js_is_j s_is_j = s_js_i s_js_is_js_i$.
They may be viewed as 
\begin{equation}\label{szk2}
\begin{split}
\text{basic operators}  &\leftrightarrow  \text{generators in $X_n$},
\\
\text{Yang-Baxter type equation} & \leftrightarrow
\text{Coxeter relation in $X_{n}$}.
\end{split}
\end{equation}

After reviewing the type A case in Section \ref{Sec:A} and BC cases in Section \ref{sec:BC},
we treat $F_4$ in Section \ref{sec:F}. 
Theorem \ref{F4:th:24} is new. It is presented with some details which were not included in \cite[Sec.4]{KO12}. 
A further generalization to the non-crystallographic Coxeter group $H_3$ is given in Section \ref{sec:H}.
Sections \ref{Sec:A}, \ref{sec:BC} and \ref{sec:H} 
are examples of $(\ref{szk1})_{n=2}$.
On the other hand, Section \ref{sec:F} correspond to $(\ref{szk1})_{n=3}$, 
and the main interest there is 
how the $F_4$ equation is decomposed into those from $B_3, C_3 \subset F_4$.
Many details are omitted in this brief note.
A full treatment can be found in the book \cite{Ku22}.

\section{$A_2 \hookrightarrow A_3$}\label{Sec:A}
We shall exclusively consider a version of the tetrahedron equation having the form:
\begin{align}\label{cdk}
R_{124}R_{135}R_{236}R_{456} = R_{456}R_{236}R_{135}R_{124}.
\end{align}
Here $R$ is a linear operator $R \in \mathrm{End}(F _q^{\otimes 3})$ for some vector space $F_q$.
The equality (\ref{cdk}) is to hold in 
$\mathrm{End}(F_q^{\otimes 6}) = \mathrm{End}(
\overset{1}{F}_q \otimes\overset{2}{F}_q \otimes\overset{3}{F}_q \otimes\overset{4}{F}_q
 \otimes\overset{5}{F}_q \otimes\overset{6}{F}_q) $, where 
$R_{ijk}$ acts on the components 
$\overset{i}{F}_q \otimes\overset{j}{F}_q \otimes\overset{k}{F}_q$ 
as $R$ and elsewhere as the identity.\footnote{The set theoretical versions is obtained by 
replacing $F_q$ by a set and $\otimes$ by the product of sets.}
We assume $R=R^{-1}$ except in Section \ref{sec:H}.
Recall that the Yang-Baxter equation corresponds to reversing 
a triangle which is a planar object.
The tetrahedron equation (\ref{cdk}) 
 is a three dimensional analogue of it in the sense that 
 the it is similarly associated with the inversion of a tetrahedron in 3D space.
This can be seen by drawing a diagram for (\ref{cdk}). 
  
One can formally associate the operator $R$
with the relation $s_1s_2s_1 = s_2s_1s_2$ 
of the generators of the Coxeter group $A_2$.\footnote{In this note, the symbol like  $A_2$ will be used 
either to mean the classical simple Lie algebra or the Coxeter group arising as its Weyl group.}
Then the tetrahedron equation (\ref{cdk}) is a natural consequence of the 
embedding $A_2 \hookrightarrow A_3$ as a parabolic subgroup.
To explain it,  consider the reduced expression (rex) graph of the longest element of $A_3$:

\begin{figure}[H]
\begin{picture}(200,100)(80,-50)
\put(-10,0){121321}
\multiput(17,11)(2.3,2){9}{\put(0,0){.}}\put(18,-3){\line(2,-2){18}}
\put(0,5){
\put(40,20){123121} \put(72,23){\line(1,0){15}}\put(90,20){123212}\put(123,23){\line(1,0){15}}
\put(140,20){132312}
\multiput(173,23)(2.3,1.9){6}{\put(0,0){.}}\multiput(173,23)(2.3,-1.9){6}{\put(0,0){.}}
\put(190,30){132132}\put(190,10){312312}
\multiput(234,23)(-2.3,1.9){6}{{.}}\multiput(234,23)(-2.3,-1.9){6}{{.}}
\put(240,20){312132}\put(273,23){\line(1,0){15}}\put(290,20){321232}
}
\put(0,-5){
\put(40,-20){212321}\put(72,-17){\line(1,0){15}}\put(90,-20){213231}
\multiput(123,-17)(2.3,1.9){6}{\put(0,0){.}}\multiput(123,-17)(2.3,-1.9){6}{\put(0,0){.}}
\put(140,-10){213213}\put(140,-30){231231}
\multiput(184,-17)(-2.3,1.9){6}{{.}}\multiput(184,-17)(-2.3,-1.9){6}{{.}}
\put(190,-20){231213}\put(222,-17){\line(1,0){15}}\put(240,-20){232123}
\put(272,-17){\line(1,0){15}}\put(290,-20){323123}
\multiput(322,-16)(2.3,2){9}{\put(0,0){.}}
}
\put(342,9){\line(-2,2){18}}
\put(340,0){321323}
\end{picture}
\caption{The rex graph for the longest element of $A_3$.}
\label{fig:rex}
\end{figure}
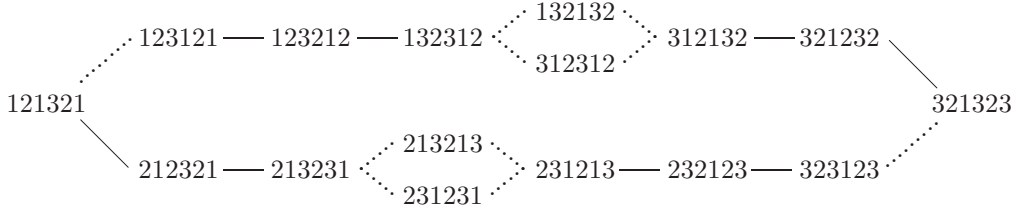

\noindent
The word $121321$ for instance means $s_1s_2s_1s_1s_3s_2s_1$ with 
$s_1,s_2,s_3$ being the generators of $A_3$.
Two reduced expressions are connected by a solid (resp. dotted) line if they are 
transformed by a single application of the cubic (resp. quadratic) Coxeter relation
$121=212$ or $232=323$ (resp. $13=31$).\footnote{The rex graph is connected \cite{Matsu64}.}
Every time they are applied, we attach an operator 
$\Phi=\Phi^{-1}  \in \mathrm{End}(F_q^{\otimes 3})$
(resp. $P=P^{-1}  \in \mathrm{End}(F_q^{\otimes 2}))$
with indices signifying the positions of the changing letters.
Here $P$ is the exchange of components
$P(u \otimes v) = v \otimes u$.
Going from $121321$ to the most distant $321323$ via the lowest path in Figure \ref{fig:rex}
gives\footnote{Indices of the operators 
referring to the {\em positions} should not be confused 
with the numbers in  Figure \ref{fig:rex} signifying the {\em labels} of the generators.}
\begin{align}\label{cdk1}
P_{34}\Phi_{123}\Phi_{345}P_{56}P_{23}\Phi_{345}\Phi_{123}.
\end{align}
Similarly the most upper path leads to
\begin{align}\label{cdk2}
\Phi_{456}\Phi_{234}P_{12}P_{45}\Phi_{234}\Phi_{456}P_{34}.
\end{align}
Let us postulate that such a composition of operators along any nontrivial loop in the rex graph yields the identity.
It amounts to setting (\ref{cdk1}) = (\ref{cdk2}).
We further relate $\Phi$ to $R$ by $\Phi_{ijk} = R_{ijk}P_{ik}$.\footnote{
$R=R^{-1}$ and $\Phi^{-1} = \Phi$ amount to assuming 
$P_{ik}R_{ijk}P_{ik} = R_{ijk}$ or equivalently $R_{ijk} = R_{kji}$.}
Substitute it into (\ref{cdk1}) = (\ref{cdk2}) and send all the $P_{ij}$' s to the right by
using $P_{34}R_{123} = R_{124}P_{34}$ etc.
The result reads 
$R_{124}R_{135}R_{236}R_{456}\sigma = R_{456}R_{236}R_{135}R_{124}\sigma'$
with 
$\sigma = P_{34}P_{13}P_{35}P_{56}P_{23}P_{35}P_{13}$
and 
$\sigma = P_{46}P_{24}P_{12}P_{45}P_{24}P_{46}P_{34}$.
Since $\sigma=\sigma'$ is the reverse ordering of the six components, (\ref{cdk}) follows.
Different choices of the initial point and the branches of the paths in the rex graph  
lead to apparently different guises  which are all equivalent to (\ref{cdk}).

The formal connection of the tetrahedron equation (\ref{cdk}) to $A_3$ 
explained so far is known to admit a concrete realization 
in the representation theory of quantized coordinate ring $A_q(A_3)$,
which leads to a solution such that $F_q$ is a 
$q$-oscillator Fock space \cite{KV94}.\footnote{The solution in \cite[eq.(30)]{BS06}
coincides with the one  in \cite[p194]{KV94} (up to typo) as shown in \cite[eq.(2.29)]{KO12}.}

\section{$B_2\hookrightarrow B_3$ and $C_2 \hookrightarrow C_3$}\label{sec:BC}

Parallel results for the quantized coordinate rings $A_q(B_3)$ and $A_q(C_3)$ have been obtained in 
\cite{KO12,KO13}. 

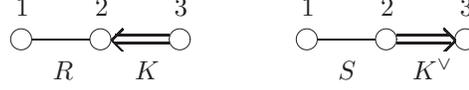
\begin{figure}[H]
\begin{picture}(250,26)(-25,-3)

\put(12,0){
\put(-2,19){1}\put(28,19){2}\put(58,19){3}
\put(12,-5){$R$}\put(43,-5){$K$}
\put(0,10){\circle{8}}\put(30,10){\circle{8}}\put(60,10){\circle{8}}
\put(4,10){\line(1,0){22}}
\put(34,10){\rotatebox{45}{\line(1,0){6}}}\put(34,10){\rotatebox{-45}{\line(1,0){6}}}

\put(56,8.5){\line(-1,0){20.5}}\put(56,11.5){\line(-1,0){20.5}}}

\put(120,0){
\put(-2,19){1}\put(28,19){2}\put(58,19){3}
\put(12,-5){$S$}\put(40,-5){$K^{\vee}$}
\put(0,10){\circle{8}}\put(30,10){\circle{8}}\put(60,10){\circle{8}}
\put(4,10){\line(1,0){22}}
\put(56,10){\rotatebox{45}{\line(-1,0){6}}}
\put(56,10){\rotatebox{-45}{\line(-1,0){6}}}
\put(34,8.5){\line(1,0){20.5}}\put(34,11.5){\line(1,0){20.5}}
}

\end{picture}
\caption{Dynkin diagrams of $C_3$ (left) and $B_3$ (right). 
The operators associated with the Coxeter relations are also indicated under them.
The both $R$ and $S$ correspond to the type A cubic one, and they are simply related 
as $S=R|_{q\rightarrow q^2}$ in the concrete realization by quantized coordinate rings.
$K$ corresponds to the quartic Coxeter relation $s_2s_3s_2s_3 = s_3s_2s_3s_2$ for $C_3$.
For $B_3$, the role of the short and the long simple roots are exchanged,
hence $K^\vee_{ijkl}  = P_{il}P_{jk}K_{ijkl} P_{il}P_{jk} =  K_{lkji}$.}
\label{fig:dynkin}
\end{figure}

We formally attach the operators 
$\Phi = \Phi_{123}$, $\Phi' = \Phi'_{123}$, 
$\Psi=\Psi_{1234}$ and $\Psi'=\Psi'_{1234}$ to the transformations 
of the subword in the reduced expressions as follows\footnote{For $K$ one needs to specify the direction of the 
map as opposed to the cubic Coxeter relation because 
the two sides are not invariant under the reverse ordering.}:
\begin{align}
C_3: \;&\Phi = R P_{13} : 121 \leftrightarrow 212,\qquad \,\Psi = KP_{14}P_{23}: 2323 \rightarrow 3232,
\label{tok1}\\
B_3: \;&\Phi' = S P_{13} : 121 \leftrightarrow 212,\qquad 
\Psi' = K^\vee P_{14}P_{23}: 2323 \rightarrow 3232.
\label{tok2}
\end{align}
Since the role of the short and the long simple roots are exchanged between $B_2$ and $C_2$,
$K$ and $K^\vee$ are related by  $K^\vee_{ijkl}  = P_{il}P_{jk}K_{ijkl} P_{il}P_{jk} =  K_{lkji}$.
We assume $K=K^{-1}$ in what follows.
It formally implies $\Psi' = \Psi^{-1}$.
We also attach $P$ to $13 \leftrightarrow 31$ for the both of $B_3$ and $C_3$.

The rex graphs 
for the longest element of $B_3$ and $C_3$ are identical and consist of 42 reduced expressions.
An example is $123121323$ in terms of indices.
Demanding again that the compositions of 
$P, \Phi, \Phi', \Psi, \Psi'$ along nontrivial loops in the rex graph 
becomes the identity, we get 
\begin{align}
C_3:\; \; &{R}_{689}{K}_{3579}{R}_{249}{R}_{258}
{K}_{1478}{K}_{1236}{R}_{456}
=
{R}_{456}{K}_{1236}
{K}_{1478}
{R}_{258}{R}_{249}
{K}_{3579}{R}_{689}
\label{F4:rk}\\
& \in \mathrm{End}\bigl(
\overset{1}{{F}}_{q^2}
\otimes \overset{2}{{F}}_{q}\otimes \overset{3}{{F}}_{q^2}
\otimes \overset{4}{{F}}_{q}\otimes \overset{5}{{F}}_{q}
\otimes \overset{6}{{F}}_{q}\otimes \overset{7}{{F}}_{q^2}
\otimes \overset{8}{{F}}_{q}\otimes \overset{9}{{F}}_{q}\bigr),
\nonumber \\
B_3:\; \; &{S}_{689}{K}_{9753}{S}_{249}{S}_{258}{K}_{8741}
{K}_{6321}{S}_{456}
= {S}_{456}{K}_{6321}{K}_{8741}{S}_{258}
{S}_{249}{K}_{9753}{S}_{689}
\label{F4:sk}\\
&\in \mathrm{End}\bigl(
\overset{1}{{F}}_{q}
\otimes \overset{2}{{F}}_{q^2}\otimes \overset{3}{{F}}_{q}
\otimes \overset{4}{{F}}_{q^2}\otimes \overset{5}{{F}}_{q^2}
\otimes \overset{6}{{F}}_{q^2}\otimes \overset{7}{{F}}_{q}
\otimes \overset{8}{{F}}_{q^2}\otimes \overset{9}{{F}}_{q^2}\bigr).
\nonumber
\end{align}
They are called 3D reflection equations and represent a ``factorization  of the three string scattering 
amplitude'' in the presence of a reflecting plane.  
See also \cite{IK97}.
Solutions of (\ref{F4:rk}) and (\ref{F4:sk}) associated with the 
quantized coordinate rings $A_q(C_3)$ and $A_q(B_3)$ have been obtained in 
\cite{KO12, KO13}.
At $q=0$ they yield the set theoretical versions where 
$F_q$ and $F_{q^2}$ are replaced by $\Z_{\ge 0}$.\footnote{It also emerges by a tropical variable change
from another set theoretical version where the $R,S,K$ become birational maps.}
The both $R$ and $S$ reduce to the map on $(\Z_{\ge 0})^3$ as 
\begin{align}\label{mas}
(a,b,c) \mapsto (b+(a-c)_+,\min(a,c),b+(c-a)_+),
\end{align}
where $(x)_+ = \max(x,0)$.
Similarly $K$ becomes a map on $(\Z_{\ge 0})^4$
defined by 
\begin{equation}\label{Cq:combk}
\begin{split}
&K: (a,b,c,d) \mapsto (a',b',c',d'),
\\
&a' = x+a+b-d,\;\;
b' = c-x+d-\min(a, c + x),\\
&c' =  \min(a, c + x),\;\;
d' = b+(c+x-a)_+,\;\;x=(c-a+(d-b)_+)_+.
\end{split}
\end{equation}

Let us present an example of the set theoretical version of (\ref{F4:rk}) and (\ref{F4:sk}),
which are equalities of maps on $(\Z_{\ge 0})^9$.

\begin{figure}[H]
\begin{picture}(100,225)(50,-85)
\unitlength=0.29mm
\put(-80,0){

 \put(62,140){Type C}
 
\put(50,120){$(211202341)$}

\put(7,110){${R}_{456}\; \swarrow$}
\put(105,110){$\searrow \,{R}_{689}$}

\put(0,90){$(211020341)$} \put(110,90){$(211205314)$} 

\put(-7,75){${K}_{1236} \downarrow$}
\put(130,75){$\downarrow {K}_{3579}$}

\put(0,60){$(301021341)$} \put(110,60){$(213205118)$} 

\put(-1,45){${K}_{1478} \downarrow$}
\put(130,45){$\downarrow {R}_{249}$}

\put(0,30){$(301021341)$} \put(110,30){$(223105119)$} 

\put(-1,15){${R}_{258} \downarrow$}
\put(130,15){$\downarrow {R}_{258}$}

\put(0,0){$(321001361)$} \put(110,0){$(213115109)$} 

\put(-7,-15){${R}_{249} \downarrow$}
\put(130,-15){$\downarrow {K}_{1478}$}

\put(0,-30){$(311101360)$} \put(110,-30){$(313015119)$} 

\put(-1,-45){${K}_{3579} \downarrow$}
\put(130,-45){$\downarrow {K}_{1236}$}

\put(0,-60){$(313101164)$} \put(110,-60){$(313015119)$} 

\put(7,-75){${R}_{689}\searrow$}
\put(105,-75){$\swarrow \,{R}_{456}$}

\put(50,-90){$(313106119)$}
}

\put(150,0){

 \put(62,140){Type B}
 
\put(50,120){$(211202341)$}

\put(7,110){$S_{456}\; \swarrow$}
\put(105,110){$\searrow \,S_{689}$}

\put(0,90){$(211020341)$} \put(110,90){$(211205314)$} 

\put(-7,75){${K}_{6321} \downarrow$}
\put(130,75){$\downarrow {K}_{9753}$}

\put(0,60){$(401021341)$} \put(110,60){$(213205116)$} 

\put(-1,45){${K}_{8741} \downarrow$}
\put(130,45){$\downarrow S_{249}$}

\put(0,30){$(301021431)$} \put(110,30){$(223105117)$} 

\put(-1,15){$S_{258} \downarrow$}
\put(130,15){$\downarrow S_{258}$}

\put(0,0){$(321001451)$} \put(110,0){$(213115107)$} 

\put(-7,-15){$S_{249} \downarrow$}
\put(130,-15){$\downarrow {K}_{8741}$}

\put(0,-30){$(311101450)$} \put(110,-30){$(413015117)$} 

\put(-1,-45){${K}_{9753} \downarrow$}
\put(130,-45){$\downarrow {K}_{6321}$}

\put(0,-60){$(314101153)$} \put(110,-60){$(314014117)$} 

\put(7,-75){$S_{689}\searrow$}
\put(105,-75){$\swarrow \,S_{456}$}

\put(50,-90){$(314105117)$}
 }
\end{picture}    
\caption{
Examples of the set theoretical solution of the 3D reflection equations for type B and C on $(\Z_{\ge 0})^9$.
$R$ and $S$ are given by (\ref{mas}).
$K_{ijkl}$  is given by (\ref{Cq:combk}) if  $i<j<k<l$ and by 
$P_{il}P_{jk}(K_{lkji} \text{ by (\ref{Cq:combk})}) P_{il}P_{jk}$ if $i>j>k>l$.}
\end{figure}
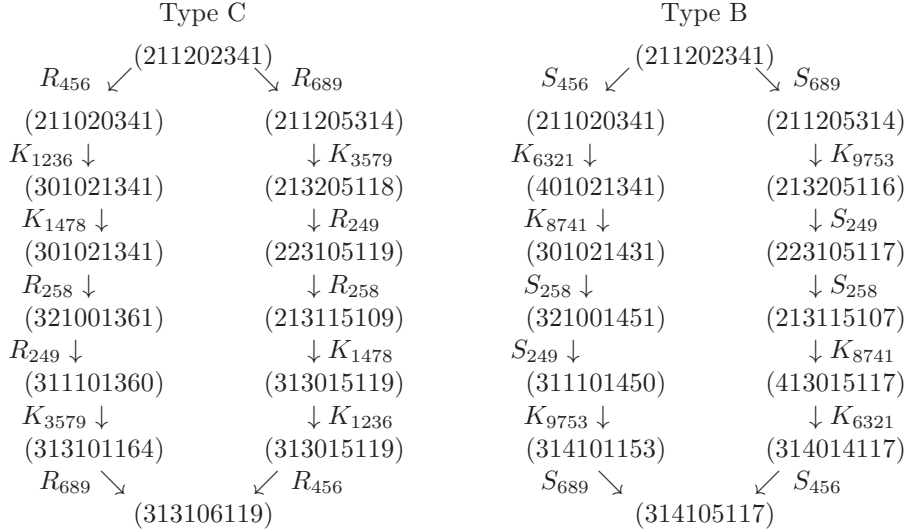

\section{$B_3 \hookrightarrow F_4 \hookleftarrow C_3$}\label{sec:F}
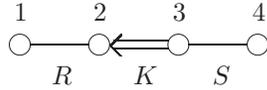
\begin{figure}[H]
\begin{picture}(250,22)(40,-3)

\put(120,0){
\put(-2,19){1}\put(28,19){2}\put(58,19){3}\put(88,19){4}
\put(12,-5){$R$}\put(43,-5){$K$}\put(73,-5){$S$}
\put(0,10){\circle{8}}\put(30,10){\circle{8}}\put(60,10){\circle{8}}\put(90,10){\circle{8}}
\put(4,10){\line(1,0){22}}\put(64,10){\line(1,0){22}}
\put(34,10){\rotatebox{45}{\line(1,0){6}}}\put(34,10){\rotatebox{-45}{\line(1,0){6}}}
\put(56,8.5){\line(-1,0){20.5}}\put(56,11.5){\line(-1,0){20.5}}
}

\end{picture}
\caption{The Dynkin diagram of $F_4$. 
The operators $R,S$ and $K$ are associated according to Figure \ref{fig:dynkin}.}
\label{fig:f4}
\end{figure}

Let us consider $F_4$ which contains $B_3$ and $C_3$ as parabolic subgroups.
Compare Figure \ref{fig:dynkin} and Figure \ref{fig:f4}.

As before we attach $P$ to the quadratic Coxeter relation $ij=ji$ 
(in terms of indices of the 
generators) with $|i-j|\ge 2$.
Furthermore in view of  (\ref{tok1}) and (\ref{tok2}), we set 
\begin{align}\label{F4:Dpsi}
\Phi & = R P_{13} : 121 \leftrightarrow 212, \quad \Upsilon  = SP_{13}: 343 \leftrightarrow 434,
\quad
\Psi  = KP_{14}P_{23}: 2323 \rightarrow 3232.
\end{align}
We have used the symbol $\Upsilon$ refreshing $\Phi'$ in (\ref{tok2}) 
since the relevant letters are now 3 and 4 instead of 1 and 2.
We keep assuming $\Phi= \Phi^{-1}, \Upsilon = \Upsilon^{-1}$
and  $R=R^{-1}, S=S^{-1}, K = K^{-1}$.
They imply
$R_{ijk}=R_{kji}, S_{ijk}=S_{kji}$ and 
$\Psi^{-1}_{ijkl} = K^\vee_{ijkl}P_{il}P_{jk}= K_{lkji}P_{il}P_{jk}$ as before.

An example of reduced expressions of the  longest element of the Coxeter group $F_4$ is
${\bf w}_0 = 434234232123423123412321$
in terms of indices.
It has length 24.
The rex graph for it consists of $2144892$ vertices.
Let  $\tilde{\bf w}_0$ be the reverse ordering of ${\bf w}_0$,
which is most distant from it in the graph.
One way to go from ${\bf w}_0$ to $\tilde{\bf w}_0$ is shown below,
where the underlines indicate the changing part and the relevant 
operators are given on the right for each step.

%{\small 
\begin{alignat}{2}
{\bf w}_0 :\;\;  &\underline{434} 23\underline{42} 3212342312\underline{341} 2321 
&\qquad&  P_{6, 7} P_{18, 19}P_{19, 20} \Upsilon_{1, 2, 3} 
  \nonumber \\ 
  &34\underline{3232} 
  432123423\underline{121} 342321 && \Psi^{-1}_{3, 4, 5, 6} 
  \Phi_{16, 17, 18} 
  \nonumber \\ &3\underline{42} 32\underline{343212} 3423212342321 
  && P_{2, 3}P_{10, 11}P_{9, 10}P_{8, 9} \Upsilon_{6, 7, 8} 
  \nonumber \\
  &3243\underline{24} 3212\underline{434} 232123\underline{42} 
  321 && P_{5, 6}P_{20, 21} \Upsilon_{11, 12, 13} 
  \nonumber \\ 
  &32\underline{434} 
  2321234\underline{32321} 2324321 && \Upsilon_{3, 4, 5}P_{16, 17} 
  \Psi^{-1}_{13, 14, 15, 16} 
  \nonumber \\ 
  &3234\underline{32321} 23\underline{42} 321
  \underline{3232} 4321 && P_{8, 9} \Psi^{-1}_{5, 6, 7, 8}P_{12, 13} 
  \Psi^{-1}_{17, 18, 19, 20} 
  \nonumber \\ 
  &323\underline{42} 321\underline{3232} 432123
  \underline{234321} && P_{4, 5} \Psi^{-1}_{9, 10, 11, 12}P_{19, 20} 
  P_{23, 24}P_{22, 23} \Upsilon_{20, 21, 22} 
  \nonumber \\ 
  &\underline{3232} 4321232
  \underline{343212} 3423214 && \Psi^{-1}_{1, 2, 3, 4}P_{16, 17}P_{15, 16} 
  P_{14, 15} \Upsilon_{12, 13, 14} 
  \nonumber \\
   &232\underline{343212} 3\underline{24} 
  3212\underline{434} 23214 && \Upsilon_{17, 18, 19}P_{11, 12}P_{8, 9} 
  P_{7, 8}P_{6, 7} \Upsilon_{4, 5, 6} 
  \nonumber \\ 
  &23243212\underline{434} 232123
   \underline{432321} 4 && \Upsilon_{9, 10, 11}P_{18, 19}P_{22, 23} 
  \Psi^{-1}_{19, 20, 21, 22} 
  \nonumber \\ 
  &23243\underline{2123432321} 232432134 && 
  P_{9, 10}P_{8, 9} \Phi_{6, 7, 8}P_{14, 15} 
  \Psi^{-1}_{11, 12, 13, 14} 
  \nonumber \\ 
  &2324\underline{31} 23412321\underline{3232} 
  432134 && P_{5, 6} \Psi^{-1}_{15, 16, 17, 18} 
  \nonumber \\ 
  &232\underline{41} 
  3234123\underline{2123} 2\underline{34321} 34 && P_{4, 5}P_{15, 16} 
  \Phi_{13, 14, 15}P_{21, 22}P_{20, 21} 
  \Upsilon_{18, 19, 20} 
  \nonumber \\ 
  &23214323412\underline{31} 23124321434 && 
  P_{12, 13} 
  \nonumber \\ 
  &232143234\underline{121} 323124321434 && 
  \Phi_{10, 11, 12} \nonumber \\ &23214323\underline{4212323} 124321434 && 
  P_{10, 11}P_{9, 10} \Psi_{12, 13, 14, 15}
   \nonumber \\ 
   &23214
  \underline{32321} 4323\underline{21243} 21434 && P_{9, 10} \Psi^{-1}_{6, 7, 8, 9}
   P_{18, 19}P_{17, 18} \Phi_{15, 16, 17} 
   \nonumber \\ 
   &2321\underline{42} 321
   \underline{3432} 3\underline{124} 3\underline{121} 434 && P_{5, 6} 
  P_{12, 13} \Upsilon_{10, 11, 12}P_{15, 16}P_{16, 17} 
  \Phi_{19, 20, 21} 
  \nonumber \\ 
  &23\underline{212} 432\underline{143} 2\underline{4341} 
  23212434 && \Phi_{3, 4, 5}P_{10, 11}P_{9, 10}P_{15, 16} 
  \Upsilon_{13, 14, 15} 
  \nonumber \\ 
  &2\underline{31} 2143\underline{24} 312\underline{34132321}
   2434 && P_{2, 3}P_{8, 9}P_{13, 14}P_{14, 15} 
  P_{19, 20} \Psi^{-1}_{16, 17, 18, 19} 
  \nonumber \\ 
  &21321\underline{434} 23
  \underline{121} 3\underline{42} 32132434 && \Upsilon_{6, 7, 8} \Phi_{11, 12, 13} 
  P_{15, 16} 
  \nonumber \\ 
    &2132\underline{13432321} 232432132434 && P_{6, 7} 
  P_{5, 6}P_{11, 12} \Psi^{-1}_{8, 9, 10, 11} 
  \nonumber \\ 
  &21323412321
  \underline{3232} 432132434 && \Psi^{-1}_{12, 13, 14, 15} 
  \nonumber \\ 
  &213234123
  \underline{2123} 2\underline{34321} 32434 && P_{12, 13} \Phi_{10, 11, 12} 
  P_{18, 19}P_{17, 18} \Upsilon_{15, 16, 17} 
  \nonumber \\ 
  &21323412\underline{31} 
  23124321432434 && P_{9, 10} 
  \nonumber \\ 
  &213234\underline{121} 
  323124321432434 && \Phi_{7, 8, 9} 
  \nonumber \\ 
  &21323\underline{4212323} 
  124321432434 && P_{7, 8}P_{6, 7} \Psi_{9, 10, 11, 12} 
  \nonumber \\ 
  &21
  \underline{32321} 4323\underline{21243} 21432434 && P_{6, 7} 
  \Psi^{-1}_{3, 4, 5, 6}P_{15, 16}P_{14, 15} 
  \Phi_{12, 13, 14} 
  \nonumber \\ 
  &\underline{2123} 21\underline{3432} 3\underline{124} 3
  \underline{121} 432434 && P_{3, 4} \Phi_{1, 2, 3}P_{9, 10} 
  \Upsilon_{7, 8, 9}P_{12, 13}P_{13, 14} \Phi_{16, 17, 18} 
  \nonumber \\ 
  &123
  \underline{1214} 32\underline{4341} 23212432434 && P_{6, 7} \Phi_{4, 5, 6} 
  P_{12, 13} \Upsilon_{10, 11, 12} 
  \nonumber \\ 
  &123214\underline{23234} 1
  \underline{32321} 2432434 && P_{10, 11} \Psi_{7, 8, 9, 10}P_{16, 17} 
  \Psi^{-1}_{13, 14, 15, 16} 
  \nonumber \\ 
  &12321432\underline{342123} 2132432434 && 
  P_{9, 10}P_{10, 11}P_{13, 14} 
  \Phi_{11, 12, 13} 
  \nonumber\\ 
  &1232143213\underline{42} 3\underline{121} 32432434 &&
  P_{11, 12} \Phi_{14, 15, 16}
  \nonumber\\
  \tilde{\bf w}_0: \;\;& 123214321324321232432434.  &&
  \label{F4:1way}
  \end{alignat}
%}
%

Let us write (\ref{F4:1way}) schematically as
\begin{align}\label{F4:15way}
{\bf w}_0 \overset{O_1}{\longrightarrow} {\bf w}_1 \overset{O_2}{\longrightarrow}
\cdots \overset{O_{N-1}}{\longrightarrow} {\bf w}_{N-1} 
\overset{O_N}{\longrightarrow}{\bf w}_N = \tilde{\bf w}_0,
\end{align} 
where $O_m\in \{P_{k,k+1}, \Phi_{k,k+1,k+2}, \Psi^{\pm 1}_{k,k+1,k+2,k+3},
\Upsilon_{k,k+1,k+2,k+3}\}$ and $N=126$.
For instance $O_1 = \Upsilon_{1,2,3}$ and 
$O_{126} = P_{11,12}$.
Considering the inverse procedure reversing the length 24 arrays at every stage, 
one finds another route going from ${\bf w}_0$ to $\tilde{\bf w}_0$ as
\begin{align}\label{F4:2way}
{\bf w}_0 = \tilde{\bf w}_N \overset{\tilde{O}_N^{-1}}{\longrightarrow} 
\tilde{\bf w}_{N-1} \overset{\tilde{O}_{N-1}^{-1}}{\longrightarrow}
\cdots \overset{\tilde{O}_2^{-1}}{\longrightarrow} \tilde{\bf w}_1 
\overset{\tilde{O}_1^{-1}}{\longrightarrow}\tilde{\bf w}_0,
\end{align} 
where $\tilde{\bf w}_r$ denotes the reverse word of ${\bf w}_r$.
The operators $\tilde{O}_m$ is chosen according to $O_m$ as  
\begin{equation}
\begin{split}
O_m  : &\;  P_{k,k+1}, \quad \Phi_{k,k+1,k+2},\quad \Upsilon_{k,k+1,k+2},
\quad \Psi^{\pm 1}_{k,k+1,k+2,k+3},
\\
\tilde{O}_m : & \; P_{j+1,j+2}, \;\; \Phi_{j,j+1,j+2},\quad \; \Upsilon_{j,j+1,j+2},
\quad \Psi^{\mp 1}_{j-1,j,j+1,j+2}
\end{split}
\end{equation}
with $j+k=23$. 
The reason for exceptionally inverting $\Psi$ is that 
the reverse ordering of $2323$ into $3232$ interchanges the role of two sides in (\ref{F4:Dpsi}).
As in the preceding cases of $A_3, B_3, C_3$, we define  
the $F_4$ analogue of the tetrahedron equation to be the condition that 
the composition of the operators along the nontrivial loops in the rex graph for the 
longest element is the identity:
\begin{align}\label{F4:f4eq1}
O_N \cdots O_2 O_1 = \tilde{O}^{-1}_1 \tilde{O}^{-1}_2 \dots \tilde{O}^{-1}_N.
\end{align}

From (\ref{F4:1way}) one obtains, after cancelling the product of $P_{i,j}$'s, the following equation:
\begin{align}
&R_{14, 15, 16} R_{9, 11, 16} K_{7, 8, 10, 16} K_{17, 15, 13, 
  9} R_{4, 5, 16} S_{7, 12, 17} R_{1, 2, 16} S_{6, 10, 
  17} R_{9, 14, 18} 
  \nonumber\\
  &\times K_{17, 5, 3, 1} R_{11, 15, 18} K_{6, 8, 
  12, 18} R_{1, 4, 18} R_{1, 8, 15} S_{7, 13, 19} K_{19, 11, 
  6, 1} K_{19, 15, 12, 4} S_{3, 10, 19} 
   \nonumber\\
  & \times R_{4, 8, 11} K_{20, 
  14, 7, 1} R_{2, 5, 18} S_{6, 13, 20} S_{3, 12, 20} R_{1, 9, 
  21} K_{20, 15, 10, 2} R_{4, 14, 21} K_{3, 8, 13, 21}
  \nonumber\\
  & \times 
   R_{2, 11, 21} R_{2, 8, 14} S_{6, 7, 22} K_{22, 4, 3, 2} R_{5, 15, 
  21} K_{22, 14, 13, 11} S_{10, 12, 22} K_{23, 9, 6, 2} S_{3, 
  7, 23}
  \nonumber\\
  & \times 
   S_{19, 20, 22} K_{22, 18, 17, 16} S_{10, 13, 
  23} K_{23, 14, 12, 5} S_{3, 6, 24} K_{23, 21, 19, 16} K_{24,
   9, 7, 4} S_{17, 20, 23} 
    \nonumber\\
   & \times
   K_{24, 11, 10, 5} S_{12, 13, 
  24} S_{17, 19, 24} K_{24, 21, 20, 18} R_{5, 8, 9} S_{22, 23, 24}
  \nonumber\\
   = & \; \text{product in reverse order},
 \label{F4:jpa}
\end{align}   
where the reverse ordering does not change the indices of $K_{i,j,k,l}$  internally into $K_{l,k,j,i}$.
There are 50 operators in total on each side; 16 $R$'s, 16 $S$'s and 18 $K$'s.
They all have distinct set of indices.

Given a reduced expression of the longest element ${\bf w}_0 = i_1 \ldots i_{24}$, one can get another one by 
${\bf w}'_0 = (5-i_1)\ldots (5-i_{24})$.
Suppose the $F_4$ analogue of the tetrahedron equation for  ${\bf w}_0$ 
is  $Z_1\cdots Z_{50} = Z_{50}\cdots Z_1$ where 
 $Z_r$ is one of $R_{ijk}, S_{ijk}$ and $K_{ijkl}$ for some $i,j,k,l \in \{1,\ldots, 24\}$.
 Then the equation for  ${\bf w}'_0$ takes the form
 $Z'_1\cdots Z'_{50} = Z'_{50}\cdots Z'_1$, where 
 $R'_{ijk} = S_{ijk}$, $S'_{ijk} = R_{ijk}$ and  $K'_{ijkl} = K_{lkji}$.
 The $F_4$ analogue of the tetrahedron equation which appeared first in \cite[eq.(48)]{KO12}
 is related to (\ref{F4:jpa}) by this transformation.
 
In Figure \ref{fig:f4} one observes the mixture of the $B_3$ and $C_3$ structures
in Figure \ref{fig:dynkin}.
This is made precise in 
 
 \begin{theorem}\cite[Th.7.2]{Ku22}\label{F4:th:24}
The $F_4$ analogue of the tetrahedron equation (\ref{F4:jpa}) is 
reduced to a composition of the 3D reflection equations for 
$B_3$ (\ref{F4:sk}) and  $C_3$ (\ref{F4:rk})
twelve times for each.
\end{theorem}

\begin{proof}
Let $X_0$ denote the expression in the  LHS of (\ref{F4:jpa}) which consists of 
16 $R$'s, 16 $S$'s and 18 $K$'s.
It can be transformed to the RHS  along the following steps:
\begin{align*}
&X_0 \rightarrow Y_0 \rightarrow X_1 \rightarrow Y_1 \rightarrow \cdots \rightarrow 
X_{24} \rightarrow Y_{24} = \text{reverse ordering of $X_0$}.
\end{align*}
Here rewriting $X_i \rightarrow Y_i$ only uses
trivial commutativity of operators having totally distinct indices.
On the other hand, the step $Y_i\rightarrow X_{i+1}$ indicates an application of 
a 3D reflection equation, which reverses seven consecutive factors somewhere 
in the length 50 array $Y_i$.
Let us label the 50 operators in $X_0$ with $1,2,\ldots, 50$  
by saying that $X_0 =1\cdot 2\cdot \cdots \cdot 50$.
Thus for instance $1=R_{14,15,16}$, 
$2 = R_{9,11,16}$, $3= K_{7,8,10,16}$ and $50=S_{22,23,24}$.
To save the space, we specify a length 50 array in two rows.
Thus $X_0$ is expressed as
$\left( {1, 2, 3, 4, 5, 6, 7, 8, 9, 10, 11, 12, 13, 14, 15, 16, 17, 18, 
  19, 20, 21, 22, 23, 24, 25 \atop 26, 27, 28, 29, 30, 31, 32, 33, 
  34, 35, 36, 37, 38, 39, 40, 41, 42, 43, 44, 45, 46, 47, 48, 49, 50}\right)$.
The intermediate forms $Y_0, Y_1, \ldots, Y_{23}$ are listed below in such a notation.\footnote{Thus 
the first one $Y_0$ already differs from $X_0$ slightly.}
\begin{small}
\begin{align*}
%1
&\left( {1, 2, 3, 4, 5, 6, 7, 8, 9, 10, 11, 12, 13, 14, 15, 16, 17, 18, 
  19, 20, 21, 22, 23, 24, 25 \atop 26, 27, 28, 29, 30, 31, 32, 33, 
  34, 35, 36, 39, 40, 41, 43, 45, 46, \blue{37}, \blue{38}, \blue{42}, \blue{44}, 
  \blue{47}, \blue{48}, \blue{50}, 49} \right) \\ 
  %2
  &\left( {1, 2, 3, 4, 5, 6, 7, 8, 9, 10, 
  11, 12, 13, 14, 15, 16, 17, 18, 19, 20, 21, 22, 23, 24, 
  25 \atop 26, 27, 28, 29, 30, 31, 32, 35, 36, 41, 43, \blue{33}, 
  \blue{34}, \blue{39}, \blue{40}, \blue{45}, \blue{46}, \blue{50}, 48, 47, 44, 42, 38, 37, 
  49}\right)\\ 
  %3
  &\left( {1, 2, 3, 4, 5, 6, 7, 8, 9, 10, 11, 12, 13, 14, 15, 16,
   17, 18, 19, 20, 21, 22, 23, 24, 25 \atop 26, 27, 28, 29, 32, 
  \blue{30}, \blue{31}, \blue{35}, \blue{36}, \blue{41}, \blue{43}, \blue{50}, 46, 45, 40, 49, 39,
   34, 33, 48, 47, 44, 42, 38, 37}\right)\\ 
   %4
   &\left( {50, 1, 2, 3, 4, 5, 6, 15,
   7, 8, 9, 10, \red{11}, \red{12}, \red{13}, \red{14}, \red{16}, \red{17}, \red{19}, 18, 
  20, 21, 22, 23, 24 \atop 25, 26, 27, 28, 29, 43, 41, 36, 35, 31, 
  30, 46, 32, 45, 40, 49, 39, 34, 33, 48, 47, 44, 42, 38, 
  37}\right)\\ 
  %5
  &\left( {50, 1, 2, 3, 5, 19, 4, 6, 15, 7, 8, 17, 10, 16, 14, 9,
   13, 20, 24, 26, 43, 18, 12, 22, 23 \atop 27, 41, 46, 36, \red{11}, 
  \red{21}, \red{25}, \red{28}, \red{32}, \red{45}, \red{48}, 29, 35, 40, 49, 47, 39, 
  44, 42, 31, 30, 34, 33, 38, 37}\right)\\ 
  %6
  &\left( {50, 1, 2, 3, 5, 19, 4, 6, 
  15, 7, 8, 17, 10, 16, 14, 9, 13, 20, 24, 26, 43, 18, \blue{12}, \blue{22}, 
  \blue{23} \atop \blue{27}, \blue{41}, \blue{46}, \blue{48}, 36, 45, 47, 32, 28, 25, 
  21, 29, 35, 40, 49, 39, 44, 42, 31, 30, 34, 11, 33, 38, 
  37}\right)\\ 
  %7
  &\left( {50, 1, 2, 3, 5, 19, 4, 6, 15, 7, 8, 17, 10, 16, 14, 
  \red{9}, \red{13}, \red{20}, \red{24}, \red{26}, \red{43}, \red{48}, 46, 18, 
  41 \atop 27, 23, 22, 36, 45, 47, 32, 28, 25, 12, 21, 29, 35, 40, 
  49, 39, 44, 42, 31, 30, 34, 11, 33, 38, 37}\right)\\ 
  %8
  &\left( \;{50, 48, 1, 2, 
  3, 5, 19, 4, 6, 15, 17, 43, 46, 7, 8, 10, 16, 14, 26, 24, 20, 13, 
  18, 41, 45 \atop 27, 23, 22, 36, 47, 32, 28, 25, \red{9}, \red{12}, 
  \red{21}, \red{29}, \red{35}, \red{40}, \red{49}, 39, 44, 42, 31, 30, 34, 11, 33, 
  38, 37}\,\right)\\ 
  %9
  &\left( \,{50, 48, 1, 2, 3, 5, 19, 4, 6, 15, 17, 43, 26, 46, 
  7, \blue{8}, \blue{10}, \blue{16}, \blue{18}, \blue{41}, \blue{45}, \blue{47}, 14, 24, 
  27 \atop 32, 49, 20, 23, 36, 40, 28, 22, 25, 35, 39, 29, 13, 21, 
  44, 42, 31, 12, 30, 34, 9, 11, 33, 38, 37}\,\right)\\ 
  %10
  &\left( \;\,{50, 48, 1, 2, 
  3, 5, 19, \blue{4}, \blue{6}, \blue{15}, \blue{17}, \blue{43}, \blue{46}, \blue{47}, 26, 45, 41,
   18, 7, 16, 10, 14, 24, 27, 32 \atop 49, 20, 23, 36, 40, 28, 8, 
  22, 25, 35, 39, 44, 29, 13, 21, 42, 31, 12, 30, 34, 9, 11, 33, 38, 
  37}\;\,\right)\\ 
  %11
  &\left(\; \;{50, 48, 47, 46, 1, 2, 3, 5, 19, 43, 45, 41, 17, 15, 6,
   26, 18, 7, 16, \red{4}, \red{10}, \red{14}, \red{24}, \red{27}, 
  \red{32} \atop \red{49}, 20, 23, 36, 40, 28, 8, 22, 25, 35, 39, 44, 29,
   13, 21, 42, 31, 12, 30, 34, 9, 11, 33, 38, 37}\;\right)\\ 
   %12
   &\left( \;\;{50, 48, 
  47, 46, 1, 2, 3, 5, 19, 43, 45, 49, 41, 17, 15, 6, 26, 18, 7, 16, 
  32, 27, 24, 14, 10 \atop 20, 23, 36, 40, 28, \blue{4}, \blue{8}, \blue{22}, 
  \blue{25}, \blue{35}, \blue{39}, \blue{44}, 29, 13, 21, 42, 31, 12, 30, 34, 9, 11, 
  33, 38, 37}\;\;\,\right)\\ 
  %13
  &\left( \;\;\,{50, 48, 47, 46, 1, 2, 3, 5, 19, 43, 45, 49, 
  41, 17, 15, 26, 18, 7, 16, 32, 27, 24, 14, \blue{6}, 
  \blue{10} \atop \blue{20}, \blue{23}, \blue{36}, \blue{40}, \blue{44}, 39, 28, 35, 42, 
  25, 22, 29, 13, 21, 31, 8, 12, 30, 34, 4, 9, 11, 33, 38, 
  37}\;\;\right)\\ 
  %14
  &\left(\;\; \;\,{50, 48, 47, 44, 46, 1, \red{2}, \red{3}, \red{5}, \red{19}, \red{43},
   \red{45}, \red{49}, 41, 17, 15, 26, 18, 7, 16, 32, 27, 24, 14, 
  40 \atop 36, 23, 20, 10, 6, 39, 28, 35, 42, 25, 22, 29, 13, 21, 
  31, 8, 12, 30, 34, 4, 9, 11, 33, 38, 37}\;\;\;\right)\\ 
  %15
  &\left(\;\; \;{50, 49, 48, 47, 
  46, 45, 44, 43, 41, 19, 1, 5, 17, 3, 15, 26, 32, 18, 40, 27, 36, 39,
   \red{2}, \red{7}, \red{16} \atop \red{24}, \red{28}, \red{35}, \red{42}, 14, 23, 20, 
  25, 22, 29, 10, 13, 21, 31, 6, 8, 12, 30, 34, 4, 9, 11, 33, 38, 
  37}\;\;\;\,\right)\\ 
  %16
  &\left( \;\;\,{50, 49, 48, 47, 46, 45, 44, 43, 41, 19, 1, 5, 17, 26, 
  32, 40, \blue{3}, \blue{15}, \blue{18}, \blue{27}, \blue{36}, \blue{39}, \blue{42}, 35, 
  28 \atop 24, 16, 7, 14, 23, 20, 25, 22, 29, 10, 13, 21, 31, 6, 8,
   12, 30, 34, 2, 4, 9, 11, 33, 38, 37}\;\;\,\right)\\ 
   %17
   &\left(\;\, {50, 49, 48, 47, 46, 
  45, 44, 43, 41, 19, \red{1}, \red{5}, \red{17}, \red{26}, \red{32}, \red{40}, \red{42}, 
  39, 36, 27, 18, 15, 35, 28, 24 \atop 16, 3, 7, 14, 23, 20, 25, 
  22, 29, 10, 13, 21, 31, 6, 8, 12, 30, 34, 2, 4, 9, 11, 33, 38, 
  37}\;\;\right)\\ 
  %18
  &\left( {50, 49, 48, 47, 46, 45, 44, 42, 40, 39, 43, 41, 36, 
  19, 32, 26, 35, 27, 28, 24, 17, 18, 23, 15, 16 \atop 5, \red{1}, 
  \red{3}, \red{7}, \red{14}, \red{20}, \red{25}, \red{29}, 22, 10, 13, 21, 31, 6, 8, 
  12, 30, 34, 2, 4, 9, 11, 33, 38, 37}\right)\\ 
  %19
  &\left( {50, 49, 48, 47, 46, 
  45, 44, 42, 40, 39, 43, 41, 36, 19, 32, 26, 35, 27, 28, 24, 17, 18, 
  23, 15, 16 \atop 5, 29, 25, 20, 14, 7, 3, 22, 10, 13, 21, 31, 6, 
  8, 12, 30, 34, \red{1}, \red{2}, \red{4}, \red{9}, \red{11}, \red{33}, \red{38}, 
  37}\right)\\ 
  %20
  &\left( {50, 49, 48, 47, 46, 45, 44, 42, 40, 39, 43, 41, 36, 
  19, 32, 26, 35, 27, 28, 24, 17, 18, 23, 29, 25 \atop 15, 16, 20, 
  14, 5, 7, 22, 10, 13, 21, 31, \blue{3}, \blue{6}, \blue{8}, \blue{12}, \blue{30}, \blue{34},
   \blue{38}, 33, 37, 11, 9, 4, 2, 1}\right)\\ 
   %21
   &\left( {50, 49, 48, 47, 46, 45, 
  44, 42, 40, 39, 43, 41, 36, 19, 32, 26, 35, 27, 28, 24, 17, 18, 23, 
  29, 25 \atop 15, 16, 20, 22, 14, \red{5}, \red{7}, \red{10}, \red{13}, \red{21},
   \red{31}, \red{38}, 34, 30, 33, 37, 12, 11, 8, 6, 3, 9, 4, 2, 
  1}\right)\\ 
  %22
  &\left( {50, 49, 48, 47, 46, 45, 44, 42, 38, 40, 39, 43, 41, 36,
   19, 32, 26, 35, 27, 28, 17, 18, 23, 29, 25 \atop 31, 34, 24, 
  \blue{15}, \blue{16}, \blue{20}, \blue{22}, \blue{30}, \blue{33}, \blue{37}, 21, 14, 13, 12, 11,
   9, 10, 8, 7, 5, 6, 3, 4, 2, 1}\right)\\ 
   %23
   &\left( {50, 49, 48, 47, 46, 45, 
  44, 42, 38, 40, 39, 43, 41, 36, 19, 32, 26, 35, 27, 28, 29, \blue{17}, 
  \blue{18}, \blue{23}, \blue{25} \atop \blue{31}, \blue{34}, \blue{37}, 33, 30, 22, 24, 
  20, 16, 15, 21, 14, 13, 12, 11, 9, 10, 8, 7, 5, 6, 3, 4, 2, 
  1}\right)\\ 
  %24
  &\left( {50, 49, 48, 47, 46, 45, 43, 41, 44, 42, 40, 39, 38, 37,
   34, 36, 35, 32, \red{19}, \red{26}, \red{27}, \red{28}, \red{29}, \red{31}, 
  \red{33} \atop 25, 23, 18, 30, 22, 24, 20, 17, 16, 15, 21, 14, 13, 
  12, 11, 9, 10, 8, 7, 5, 6, 3, 4, 2, 1}\right)
  \end{align*}
\end{small}
The blue (resp. red)\footnote{Even if not visible, they can be distinguished as explained below.} 
neighboring seven numbers  specify the place and operators 
to which the $B_3$ (resp. $C_3$) reflection equation is applied.
For example 
$X_1$ is obtained from $Y_0$ by 
replacing $37\cdot 38 \cdot  42 \cdot  44 \cdot 47 \cdot  48 \cdot 50$
$=S_{{19, 20, 22}}K_{{22, 18, 17, 16}}K_{{23, 21, 19, 16}}S_{{17, 20, 
  23}}S_{{17, 19, 24}}K_{{24, 21, 20, 18}}S_{{22, 23, 24}}$
with the reverse ordered form
$S_{{22, 23, 24}}K_{{24, 21, 20, 18}}S_{{17, 19, 24}}S_{{17, 20, 
  23}}K_{{23, 21, 19, 16}}K_{{22, 18, 17, 16}}S_{{19, 20, 22}}
  = 50 \cdot 48 \cdot 47 \cdot 44 \cdot 42 \cdot 38 \cdot 37$ 
by (\ref{F4:sk}).
The numbers of red and blue sequences are both twelve.
\end{proof}

Theorem \ref{F4:th:24} confirms that the triad 
$(R,S,K)$ satisfying the 3D reflection equations also yield 
a solution to the $F_4$ analogue of the tetrahedron  equation
(\ref{F4:jpa}).
We remark that the tetrahedron equations $RRRR=RRRR$ and $SSSS = SSSS$ have not been used.
$R$ and $S$ act as catalysts for the main reactions which are 3D reflection equations
(\ref{F4:rk}) and (\ref{F4:sk}) involving $K$.
According to \cite[Th.(2.17)]{R09},  
for any element of a Coxeter group,  
loops in its rex graph are generated 
by the loops in the rex graph of the longest element in
finite parabolic subgroups of rank 3. 
Theorem \ref{F4:th:24} is consistent with it and provides a finer information distinguishing 
$B_3$ and $C_3$  structures within $F_4$.

An analogue of Theorem \ref{F4:th:24} for $A_3 \hookrightarrow A_4$ can be found 
in \cite[eq.(3.101)]{Ku22}.

\section{$H_2 \hookrightarrow H_3$}\label{sec:H}

\begin{figure}[H]
\begin{picture}(100,15)(-10,-5)

\put(-70,0){
\put(13,6){5}
\put(0,0){$\bullet$}\put(0,-8){1}
\put(2,2){\line(1,0){25}}
\put(25,0){$\bullet$}\put(25,-8){2}
}

\put(13,6){5}
\put(0,0){$\bullet$}\put(0,-8){1}
\put(2,2){\line(1,0){50}}
\put(25,0){$\bullet$}\put(25,-8){2}
\put(50,0){$\bullet$}\put(50,-8){3}

\put(90,0){
\put(13,6){5}
\put(0,0){$\bullet$}\put(0,-8){1}
\put(2,2){\line(1,0){75}}
\put(25,0){$\bullet$}\put(25,-8){2}
\put(50,0){$\bullet$}\put(50,-8){3}
\put(75,0){$\bullet$}\put(75,-8){4}
}

\end{picture}
\caption{The Coxeter diagrams for non-crystallographic Coxeter groups $H_2, H_3$ and $H_4$.
$H_2$ is customarily denoted also by $I_2(5)$, which is the 
$m=5$ case of the dihedral groups $I_2(m)\, (m \ge 3)$.}
\label{F4:fig:H}
\end{figure}
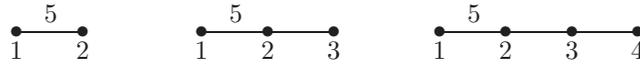

This section is a supplement concerning the non-crystallographic Coxeter groups $H_2, H_3, H_4$
and presents the $H_3$ analogue of the tetrahedron equation. 
Although there is no associated quantized coordinate ring, 
they are treated formally in a similar manner to the preceding cases of
$A_3, B_3, C_3$ and $F_4$.
The Coxeter diagrams of $H_2, H_3, H_4$ are given in Figure \ref{F4:fig:H}.

They indicate that $H_n$ is generated by $s_1,s_2,\ldots, s_n$ obeying the relations
$s_i^2=1\,(1\le i \le n)$, $(s_1s_2)^5=1$, $(s_is_{i+1})^3=1,(1<i<n)$ and 
$(s_is_j)^2=1\,(|i-j|>1)$.
The groups $H_2\subset H_3 \subset H_4$ are of order 10,120, 14400 with 
longest elements of length 5,15, 60, respectively.
$H_3$ is known as 
the symmetry of the icosahedron or equivalently the dual dodecahedron \cite{Hum90}.
The relations of the generators $s_1, s_2, s_3$ are given as
$s_1^2=s_2^2=s_3^2=1$ and 
\begin{align}\label{F4:str}
s_1s_3=s_3s_1, \quad s_1s_2s_1s_2s_1 = s_2s_1s_2s_1s_2.
\end{align}

Following the preceding examples, we attach the operators to (\ref{F4:str}) as 
\begin{align}
&P=P^{-1}: 13 \leftrightarrow 31,
\label{F4:H3P}\\
&\Phi: 232 \rightarrow 323, \quad \Phi_{ijk} = R_{ijk}P_{ik},  
\label{F4:H3R}\\
&\Omega: 21212 \rightarrow 12121, \quad \Omega_{ijklm} = Y_{ijklm}P_{im}P_{jl},
\label{F4:H3O}
\end{align}
where, as before,  the indices $i,j,k,\ldots$ specify 
the components that they act non-trivially.\footnote{Unlike the $R$ for 
$A_3, B_3, C_3, F_4$, we do not assume $R^{-1}=R$ nor $R_{ijk} = R_{kji}$ in this section.}
The operators $\Omega$ and $Y$ are the characteristic ones which emerges from $H_2$.

A reduced expression of the longest element of $H_3$ is 
$ 1 2 1 2 1 3 2 1 2 1 3 2 1 2 3$ in terms of indices of $s_i$.
Now a process analogous to (\ref{F4:1way}) reads as
\begin{alignat}{2}
& 1 2 1 2 \underline{1 3} 2 1 2 1 3 2 1 2 3 &\quad &P_{5,6}
\nonumber \\
& 1 2 1 2 3 \underline{1 2 1 2 1} 3 2 1 2 3 && \Omega^{-1}_{6,7,8,9,10}
\nonumber \\
& 1 2  1 \underline{2 3 2} 1 2 1 \underline{2 3 2} 1 2 3 && \Phi_{4,5,6}\Phi_{10,11,12}
\nonumber \\
& 1 2 \underline{1 3} 2 \underline{3 1} 2 \underline{1 3} 2 \underline{3 1} 2 3 && P_{3,4}P_{6,7}P_{9,10} P_{12,13}
\nonumber \\
& 1 2 3 1 2 1 \underline{3 2 3} 1 2 1 \underline{3 2 3} && \Phi^{-1}_{7,8,9} \Phi^{-1}_{13,14,15}
\nonumber \\
& 1 2 3 1 2 1 2 3 \underline{2 1 2 1 2} 3 2 && \Omega_{9,10,11,12,13}
\nonumber \\
& 1 2 3 1 2 1 2 \underline{3 1} 2 1 2 \underline{1 3} 2 && P_{8,9} P_{13,14}
\nonumber \\
& 1 2 3 \underline{1 2 1 2 1} 3 2 1 2 3 1 2 &&  \Omega^{-1}_{4,5,6,7,8}
\nonumber \\
& 1 \underline{2 3 2} 1 2 1 \underline{2 3 2} 1 2 3 1 2 && \Phi_{2,3,4} \Phi_{8,9,10}
\nonumber \\
& 1 3 2 \underline{3 1} 2 \underline{1 3} 2 \underline{3 1} 2 3 1 2 && P_{4,5} P_{7,8}P_{10,11}
\nonumber\\
& \underline{1 3} 2 1 \underline{3 2 3} 1 2 1 \underline{3 2 3} 1 2 && P_{12}\Phi^{-1}_{567}\Phi^{-1}_{11,12,13}
\nonumber\\
& 3 1 2 1 2 3 \underline{2 1 2 1 2} 3 2 1 2 && \Omega^{-1}_{7,8,9,10,11}
\nonumber\\
& 3 1 2 1 2 \underline{3 1} 2 1 2 \underline{1 3} 2 1 2 && P_{6,7}P_{11,12}
\nonumber\\
& 3 \underline{1 2 1 2 1} 3 2 1 2 3 1 2 1 2 && \Omega^{-1}_{2,3,4,5,6}
\nonumber\\
& 3 2 1 2 1 \underline{2 3 2} 1 2 3 1 2 1 2 && \Phi_{6,7,8}
\nonumber\\
& 3 2 1 2 \underline{1 3} 2 \underline{3 1} 2 3 1 2 1 2 && P_{5,6}P_{8,9}
\nonumber\\
& 3 2 1 2 3 1 2 1 \underline{3 2 3} 1 2 1 2 && \Phi^{-1}_{9,10,11}
\nonumber\\
& 3 2 1 2 3 1 2 1 2 3 \underline{2 1 2 1 2} && \Omega_{11,12,13,14,15}
\nonumber\\
& 3 2 1 2 3 1 2 1 2 3 1 2 1 2 1. &&
\label{F4:seqH}
\end{alignat}
It reverses the initial reduced expression.
There is another route achieving the reverse ordering in the same manner as explained in 
(\ref{F4:2way}) for $F_4$.
Equating the two ways, substituting 
(\ref{F4:H3P}) --  (\ref{F4:H3O}) and 
using 
$P_{4,7}Y_{1,3,4,9} = Y_{1,3,7,9}P_{4,7}$ etc, 
we get the $H_3$ analogue of the tetrahedron equation \cite[eq.(9.12)]{Ku22}:
\begin{equation}\label{F4:H3}
\begin{split}
&Y_{11, 12, 13, 14, 15} R^{-1}_{15, 10, 9} R_{5, 7, 15} 
Y^{-1}_{15, 6, 4, 3, 2} Y_{2, 5, 8, 10, 14} R^{-1}_{14, 7, 3} R^{-1}_{13, 9, 2} 
R_{1, 6, 14} 
   \\
   & \times R_{3, 8, 13} Y^{-1}_{13, 10, 7, 4, 1} Y_{1, 3, 5, 9, 12} 
   R^{-1}_{12, 8, 4} R^{-1}_{11, 2, 1} R_{6, 10, 12} R_{4, 5, 11} Y^{-1}_{11, 9, 8, 7, 6}
  \\
& =
Y_{6, 7, 8, 9, 11} R^{-1}_{11, 5, 4} R^{-1}_{12, 10, 6} R_{1, 2, 11} 
R_{4, 8, 12} Y^{-1}_{12, 9, 5, 3, 1} Y_{1, 4, 7, 10, 13} R^{-1}_{13, 8, 3} 
  \\
  &\times R^{-1}_{14, 6, 1} R_{2, 9, 13} R_{3, 7, 14} Y^{-1}_{14, 10, 8, 5, 2} 
  Y_{2, 3, 4, 6, 15} R^{-1}_{15, 7, 5} R_{9, 10, 15} Y^{-1}_{15, 14, 13, 12, 11}.
\end{split}
\end{equation}
The two sides have the form of the inverse of each other if 
the indices within each operator were reversed.
There are  3 $Y$'s, 3 $Y^{-1}$'s, 5 $R$'s and 5 $R^{-1}$'s on each side.

If $Y^{-1}_{ijklm}=Y_{ijklm}=Y_{mlkji}$ and $R^{-1}_{ijk}=R_{ijk}=R_{kji}$
are valid,
the above equation reduces to \cite[eq.(9.13)]{Ku22}:
\begin{equation}\label{H:H32}
\begin{split}
&Y_{11, 12, 13, 14, 15} R_{9,10,15} R_{5, 7, 15} 
Y_{2, 3, 4, 6, 15} Y_{2, 5, 8, 10, 14} R_{3, 7, 14}R_{2, 9, 13}  
R_{1, 6, 14} 
   \\
   & \times R_{3, 8, 13} Y_{1, 4, 7, 10, 13} Y_{1, 3, 5, 9, 12} 
R_{4, 8, 12}R_{1, 2, 11} R_{6, 10, 12} R_{4, 5, 11} Y_{6, 7, 8, 9, 11} 
  \\
& =\text{product in reverse order}.
\end{split}
\end{equation}
It remains a challenge to construct a solution of (\ref{F4:H3}) or the reduced version (\ref{H:H32}).
A planar graphical representation of (\ref{H:H32}) has appeared in \cite[eq.(4.9)]{EW}.
In view of $H_3, A_3  \subset  H_4$,  the $H_4$ equation 
should be decomposed into the tetrahedron equation and the $H_3$ equation similarly to 
Theorem \ref{F4:th:24}.

\section*{Acknowledgments}
The author thanks Kohei Motegi for invitation to the workshop
and Toshiyuki Tanisaki for useful communication.


\begin{thebibliography}{99}

\bibitem{Bax82}
R.~ J.~ Baxter,
\textit{Exactly solved models in statistical mechanics},
Academic Press, New York  (1982)

\bibitem{BS06}
V.~V.~Bazhanov, S.~M.~Sergeev,
Zamolodchikov's tetrahedron equation and hidden structure of quantum groups,
J. Phys. A: Math. Theor. \textbf{39} 3295--3310 (2006)

\bibitem{Ch84}
I.~V.~Cherednik,
Factorizing particles on a half-line and root systems,
Theor. Math. Phys. {\bf 61}  977--983 (1984)

\bibitem{EW}B.~Elias, G.~Williamson,
Soergel calculus, 
Represent. Theory {\bf 20} 295--374 (2016) 

\bibitem{Hum90}
J.~E.~Humphreys,
{\it Reflection groups and Coxeter groups},
Cambridge Univ. Press  (1990)

\bibitem{IK97}
A.~P.~Isaev, P.~P.~Kulish,
Tetrahedron reflection equations.
Mod. Phys. Lett. A{\bf 12} 427--437 (1997)

\bibitem{Jim90}
M.~Jimbo, eds.
{\it Yang-Baxter Equation in Integrable Systems},
Advanced Series in Mathematical Physics, 
World Scientific (1990)

\bibitem{KV94}
M.~M.~Kapranov, V.~A.~Voevodsky,
2-Categories and Zamolodchikov tetrahedron equations. 
Proc. Symposia in Pure Math. \textbf{56} 177--259 (1994)

\bibitem{Kun18-1}A.~Kuniba,
Matrix product solutions to the $G_2$ reflection equation,
J. of Integrable Syst. {\bf 3}  1-28 (2018) 

\bibitem{Ku22}A.~Kuniba,
{\it Quantum groups in three-dimensional integrability}, to appear

\bibitem{KO12} A.~Kuniba, M.~Okado,
Tetrahedron and 3D reflection equations
from quantized algebra of functions , 
J. Phys. A: Math.Theor. {\bf 45}  465206 (27pp)  (2012)  

\bibitem{KO13}A.~Kuniba, M.~Okado,
A solution of the 3D reflection equation 
from quantized algebra of functions of type $B$,
in Nankai Series in Pure, Applied Mathematics and Theoretical Physics 
\textbf{11}, p181-190, World Scientific (2013)

\bibitem{Matsu64}
H.~Matsumoto,
G\'en\'rateurs et relations des groupes de Weyl g\'en\'eralis\'es, 
C. R. Acad. Sci. Paris, {\bf 258}  3419--3422 (1964)

\bibitem{R09}M.~Ronan,
{\it Lectures on buildings, updated and revised},
University of Chicago Press, Chicago IL (2009) 

\bibitem{Zam79}
A.~B.~Zamolodchikov, Al.~B.~Zamolodchikov,
Factorized S-matrices in two-dimensions as the exact solutions of certain relativistic quantum field theory models,
Ann. Phys. {\bf 120} 253--291 (1979)


\bibitem{Zam80}
A.~B.~Zamolodchikov,
Tetrahedra equations and integrable systems in three-dimensional space,
Soviet Phys. JETP {\bf 79} 641--664 (1980)

\end{thebibliography}
\end{document}